\documentclass[12pt]{article}
\usepackage{a4wide}
\usepackage{graphicx}

\newcommand{\be}{\begin{equation}}
\newcommand{\ee}{\end{equation}}
\newcommand{\ba}{\begin{eqnarray}}
\newcommand{\ea}{\end{eqnarray}}
\newcommand{\dsp}{\displaystyle}

\begin{document}
\begin{titlepage}
\begin{flushright}
LU TP 06-11\\
hep-lat/0602019\\
Revised March 2006
\end{flushright}
\begin{center}
\vfill
{\Large\bf Finite Volume Dependence of the Quark-Antiquark
Vacuum Expectation Value}
\vfill
{\bf Johan Bijnens and Karim Ghorbani\\[1cm]
Department of Theoretical Physics, Lund University\\
S\"olvegatan 14A, SE 22362 Lund, Sweden}
\end{center}
\vfill
\begin{abstract}
A general formula is derived for 
the finite volume dependence of vacuum expectation values
 analogous
to L\"uscher's formula for the masses.
The result involves no integrals over kinematic quantities and
depends only on the matrix element between pions at zero momentum transfer
thus presenting a new way to calculate the latter, i.e. pion sigma terms.

The full order $p^6$ correction to the vacuum condensate
$\langle\bar q q\rangle$ is evaluated and compared with the result from
the L\"uscher formula. Due to the size of the $p^6$ result no
conclusion about the accuracy of the L\"uscher formula can be drawn.
\end{abstract}

\vfill
\end{titlepage}

\section{Introduction}

Quantum Chromo Dynamics (QCD) at low energy remains a difficult problem.
One of the ways to deal with this problem is to numerically evaluate
the functional integral of QCD. This approach, known as lattice QCD,
has now reached the stage where realistic calculations with fairly light
quark masses are now possible. One side effect of this is that finite
volume corrections are becoming more important. Luckily in many cases
these corrections can be evaluated analytically using Chiral Perturbation
Theory (ChPT) \cite{Weinberg,GL1}. The application of ChPT to finite
volume was started by Gasser and Leutwyler \cite{GL2}. A review of recent work
in this area can be found in \cite{Colangelo}. 
Note that ChPT is applicable to finite volume as soon as the typical momenta
that are relevant are small enough. This imposes a size restriction on the
volume as
\be
F_\pi L > 1\,.
\ee
Here $F_\pi$ is the pion decay constant and $L$ is the linear size of the
volume. This paper is concerned with the $p$-regime. This is the regime
where the volume is large enough such that the zero momentum fluctuations
of the meson fields can be treated perturbatively. This is the regime
with in addition the requirement that
\be
m_\pi^2 F_\pi^2 V >> 1\,.
\ee
These finite volume corrections
have been evaluated for many quantities up to one-loop order. 
This is
the order where the first nontrivial dependence on the volume shows up.
One purpose of this paper is to calculate the full two-loop finite volume
corrections to the vacuum condensate $\langle \bar q q\rangle$. This
is the one of the calculations of finite volume effects to this
order\footnote{Ref.~\cite{CH} with another
calculation appeared essentially simultaneously. There the pion mass
at finite volume was studied at two-loop order.}
The vacuum condensate at finite volume has been studied at one-loop
in Ref.~\cite{Descotes}.

An alternative approach to finite volume corrections was
introduced by L\"uscher where the leading part of the finite volume corrections
was derived to all orders in perturbation theory for the mass in terms of
a scattering amplitude \cite{Luscher1,Luscher2}. This was extended to
the finite volume corrections for the pion decay constant in \cite{CH1}.
The other purpose of this paper is to extend the L\"uscher formula
also to vacuum expectation values. This will in general connect the
finite volume corrections of an operator to the zero-momentum transfer
matrix element of that operator between pion states as shown in
Eq.~(\ref{resultluscher}). This allows for new ways to calculate sigma terms
from the finite volume variation of vacuum condensates.

Note that all our formulas are for the case of an infinite extension
in the time direction but a finite volume in the three spatial directions.
The formulas can be easily extended to a small fourth direction as well by
replacing the integrals with the expressions for that case.

\section{A L\"uscher Formula for the Vacuum Condensate}

It turns out to be straightforward to extend L\"uscher's formula
for the mass to the case of the vacuum condensate. In fact the
formula has an even simpler structure than the one for the mass or the
decay constant \cite{CH1}.
The underlying observation of L\"uscher's method is that the leading
finite volume corrections come from one of the propagators feeling the finite
volume effects only. We write the finite volume propagator $G_V(q)$ as
\be
G_V(q) = \sum_{\vec n} e^{-iq_i n_i L_i} G_\infty(q)\,.
\ee
Here the sum is over a vector of integers and $L_i$ is the length of the
volume in the $i$-th direction. The term with $\vec n = \vec o$ gives
the infinite volume case. L\"uscher then kept only the the term
with one of the $n_i=\pm 1$ since this is parametrically leading.
As suggested in \cite{CDH} one can also easily keep the entire series,
thus keeping part of the subleading corrections.
The remainder of the L\"uscher's analysis goes by taking the
component of $\vec q$ parallel to $\vec n$, $q_n$ and distorting the
integration contour along that direction to $q_n\to q_n-i s$.
In the limit of $s\to\infty$ the contribution from that integral vanishes
and only the parts coming from the singularities encountered while deforming
the contour remain. These happen when relevant propagators can go on-shell.
A pedagogical introduction can be found in \cite{Luscher2}.

\begin{figure}[htbp]
  \begin{center}
    \includegraphics[width=0.5\textwidth]{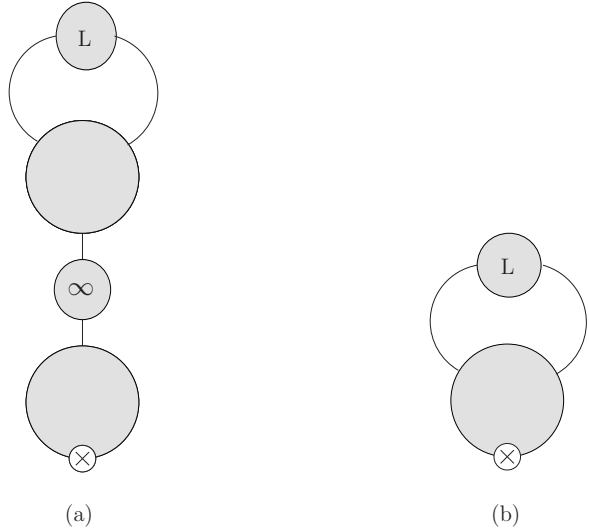}    
    \caption{The two types of diagrams giving the leading finite volume
    corrections}
    \label{figluscher}
  \end{center}
\end{figure}

The difficulty in proving this is that it needs to be done to all orders
in perturbation theory, this was done in \cite{Luscher1}. Here we only sketch
the lines of reasoning. Of the three types of contributions shown in
Fig.~4 of \cite{Luscher1} only one is relevant here and is shown in
Fig.~\ref{figluscher}(b). The case of operators with contributions of the type
shown in diagram (a) is not relevant for ChPT, one can always add the
parity-conjugate operator to remove the contribution from diagram (a).

One difference with the case of the mass or decay constant here is that
when the singularity is encountered, there is no freedom left in the
relevant matrix element and the integral over the momentum of the propagator
can be fully done. In the case of the mass and decay constant there
is an external momentum available, $p_{ext}$, this leaves after putting
the deformed $q$ on-shell a freedom in $q\cdot p_{ext}$ which results
in the final integration over $\nu$. Here there is no external momentum and
hence there is no such freedom left.

The formula after putting everything on-shell becomes
\be
\langle O \rangle_V-\langle O \rangle_\infty =
-\sum_{\vec n\ne \vec o} \frac{1}{16\pi^2}
\int_0^\infty \frac{dq^2 q^2}{\sqrt{m_0^2+q^2}} e^{-\sqrt{\vec n^2
(m_0^2+q^2)L^2}} \langle\phi| O | \phi\rangle\,,
\ee
for a real boson with infinite volume mass $m_0$ and the matrix element
on the right hand side should be taken at zero momentum transfer.
The integral can now be done explicitly in terms of the generalized Bessel
function $K_1$. It also only depends on $k=\vec n^2$ which is also integer.
The number of times in the sum that $\vec n^2 =k$ we call $x(k)$. We obtain
\be
\label{resultluscher}
\langle O \rangle_V-\langle O \rangle_\infty =
-\sum_{k=1,\infty}\frac{x(k)}{16\pi^2} \frac{m_0^2}{\sqrt{\zeta(k)}} 
K_1(\zeta(k))
 \langle\phi| O | \phi\rangle\,,
\ee
with $\zeta(k) = \sqrt{k}\, m_0 L$. Note that the above formula is for the
case of one real scalar. The multiplicity factors for complex scalars
can be trivially taken into account. Note that we have left the sum over all
modes in as suggested for the mass in~\cite{CD}.

In particular for the case of $\langle \bar q q\rangle$ the relevant
matrix element is the sigma term. The finite volume corrections to the
vacuum condensate thus are another option to calculate the pion sigma term.
This can then be compared with the direct calculation of the sigma
term via the matrix element $\langle \pi | \bar q q | \pi\rangle$
or via the Feynman-Hellman theorem from ${\partial m_\pi^2}/{\partial m_q}$.

\section{The finite volume vacuum condensate at two-loops}

\begin{figure}
\begin{center}
\includegraphics[width=0.8\textwidth]{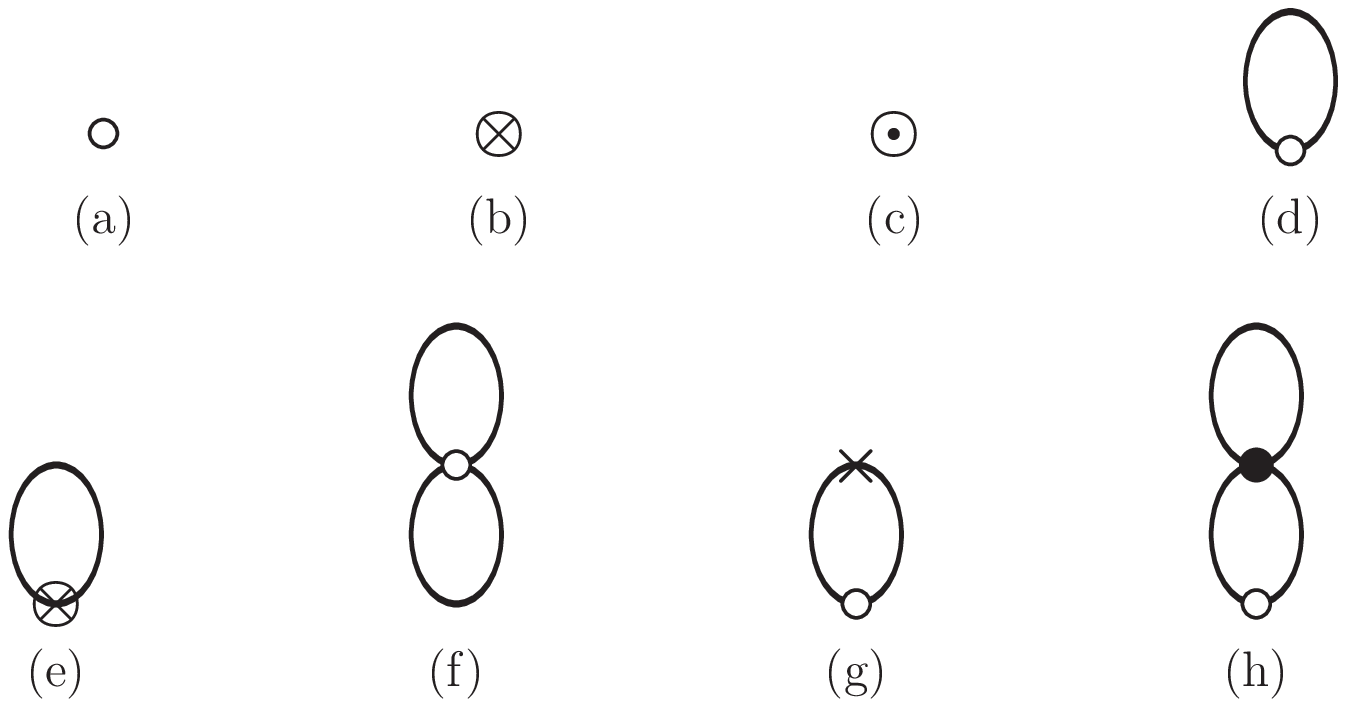}
\end{center}
\caption{\label{fig:diagrams} The diagrams up to order $p^6$ for
$\langle\overline qq\rangle$. The lines are meson propagators and the vertices
are:
$\circ$ a $p^2$ insertion of $\overline qq$,
$\otimes$ a $p^4$ insertion of $\overline qq$,
$\odot$ a $p^6$ insertion of $\overline qq$,
$\bullet$ a $p^2$ vertex and
$\times$ a $p^4$ vertex.}
\end{figure}

The vacuum condensate at two-loop was calculated in \cite{ABT1}, here we
repeat that calculation taking into account the finite volume effects.
The calculation in terms of the lowest order meson masses is straightforward
and proceeds exactly as in \cite{ABT1}. The details of calculation of two loops
in ChPT can be found in \cite{ABT1}. The diagrams that contribute are shown
in Fig.~\ref{fig:diagrams}.

In infinite volume the loop diagrams contain the integrals
\ba
A(m^2) &=& \frac{1}{i}\int \frac {d^dp}{(2\pi)^d}\,\frac{1}{p^2-m^2}\,,
\nonumber\\
B(m^2) &=&  \frac{1}{i}\int \frac {d^dp}{(2\pi)^d}\,\frac{1}{(p^2-m^2)^2}\,.
\ea
These are expanded in terms of $\epsilon = (4-d)/2$ as
\ba
A(m^2) &=& \frac{m^2}{16\pi^2}\left(\frac{1}{\epsilon}-\gamma_E+\log(4\pi) +1)
\right) + \overline A(m^2) + \epsilon A^\epsilon(m^2) + {\cal O}(\epsilon^2)\,,
\nonumber\\
B(m^2) &=& \frac{1}{16\pi^2}\left(\frac{1}{\epsilon}-\gamma_E+\log(4\pi) +1)
\right) + \overline B(m^2) + \epsilon B^\epsilon(m^2) + {\cal O}(\epsilon^2)\,,
\ea
with a similar expansion for $B(m^2)$. At finite volume the integrals over
momenta get replaced by sums over the finite possible momenta.
In this paper we only keep three of the four dimensions at the same
finite length $L$. Their principle of evaluation can be found in
Refs.~\cite{GL2,HL} and explicit expressions are at finite volume.
After renormalization the divergent parts cancel and the finite
parts become subtraction-scale $\mu$ dependent.
The needed explicit expressions are
\ba
\overline A(m^2) &=& -\frac{m^2}{16\pi^2}\log\left(\frac{m^2}{\mu^2}\right)
-\frac{1}{16\pi^2}
\sum_{k=1,\infty} x(k) \frac{4 m^2}{\lambda_k} K_1(\lambda_k)\,
\nonumber\\
\overline B(m^2) &=& -\frac{1}{16\pi^2}\log\left(\frac{m^2}{\mu^2}\right)
-\frac{1}{16\pi^2}
+\frac{1}{16\pi^2}\sum_{k=1,\infty} x(k) 2 K_0(\lambda_k)
\ea
with
$\lambda_k=\sqrt{k}\, m L$\,. The functions $K_1$ and $K_2$ are modified
Bessel functions and the integer quantity $x(k)$ is the number of times
the sum of squares $\vec k^2 = k_1^2+k_2^2+k_3^2$ is equal to $k$
when $k_1,k_2,k_3$ are varied over all positive and negative integers.
For large $L$ the Bessel functions lead to an exponential fall-off with
the finite size.

We have checked explicitly that at two-loop order the contributions
containing $A^\epsilon$ and $B^\epsilon$ cancel, so we do not need to evaluate
these at finite volume. That this cancellation happens follows from the
cancellations of nonlocal divergences but in the calculation of \cite{ABT1}
this calculation was not explicitly checked. We have also reperformed the
full calculation since in \cite{ABT1} the relation
$\overline{A}(m^2) = m^2\left( \overline{B}(m^2)+1/(16\pi^2)\right)$ was used,
which is no longer true at finite volume. 
In Ref.~\cite{ABT1} the result was expressed in terms of the physical
mass and decay constant. This can be done here as well but since we
must then distinguish between the integrals at finite volume and those
at infinite volume the formula is shorter if we leave meson masses
and decay constants in their unrenormalized form.

The result, in three flavour ChPT, uses the $p^4$ low-energy constants
(LECs) $L_i^r$ from Ref.~\cite{GL1} and the $C_i^r$ from~\cite{BCE}.
The latter drop out when comparing finite and infinite volume corrections
and the values of the $L_i^r$ only contribute to that difference
at NNLO. We also suppress the $\mu$ dependence of all quantities.
The lowest order masses for the pions, kaons and eta we denote
by
$\chi_\pi$, $\chi_K$ and $\chi_\eta$ respectively and in term
of the average up and down quark-mass $\hat m$ and strange quark-mass $m_s$
they are
\be
\chi_\pi = 2 B_0 \hat m\,,\quad \chi_K = B_0(\hat m+m_s)\,\quad
\chi_\eta = 2 B_0(\hat m+2 m_s)/3\,.
\ee

We define the quantities
\be
\langle \overline uu\rangle =
- B_0 F_0^2\left(1+\frac{v^u_4}{F_0^2}+\frac{v^u_6}{F_0^4}\right)\,,
\quad
\langle \overline ss\rangle =
- B_0 F_0^2\left(1+\frac{v^s_4}{F_0^2}+\frac{v^s_6}{F_0^4}\right)\,.
\ee
The calculation gives
\ba
v^u_4 &=&
        \overline{A}(\chi_\pi) \, \Big( 3/2 \Big)
       + \overline{A}(\chi_K) 
       + \overline{A}(\chi_\eta) \, \Big( 1/6 \Big)
       + 16\,\chi_\pi\,L^r_{6}+ 8\,\chi_\pi\,L^r_{8}
       + 4\,\chi_\pi\,H^r_{2}+ 32\,\chi_K\,L^r_{6}
  \,,
\nonumber\\
v^s_4 &=&
        \overline{A}(\chi_K) \, \Big( 2 \Big)
       + \overline{A}(\chi_\eta) \, \Big( 2/3 \Big)
       + 16\,\chi_\pi\,L^r_{6}+ 4\,(2\chi_K-\chi_\pi)\,(2\,L^r_{8}+H^r_{2})
       + 32\,\chi_K\,L^r_{6}\,,
\nonumber\\
v^u_6 &=&
        \overline{A}(\chi_\pi)^2 \, \Big(  - 3/8 \Big)
       + \overline{A}(\chi_\pi)\,\overline{A}(\chi_\eta) \, \Big( 1/4 \Big)
       + \overline{A}(\chi_\pi)\,\overline{B}(\chi_\pi) \, 
            \Big(  - 3/4\,\chi_\pi \Big)
\nonumber\\&&
       + \overline{A}(\chi_\pi)\,\overline{B}(\chi_\eta) \, \Big( 1/12\,\chi_\pi \Big)
       + \overline{A}(\chi_K)\,\overline{A}(\chi_\eta) \, \Big(  - 1/3 \Big)
       + \overline{A}(\chi_K)\,\overline{B}(\chi_\eta) \, \Big(  - 2/9\,\chi_K \Big)
\nonumber\\&&
       + \overline{A}(\chi_\eta)^2 \, \Big( 1/72 \Big)
       + \overline{A}(\chi_\eta)\,\overline{B}(\chi_\pi) \, \Big( 1/4\,\chi_\pi \Big)
       + \overline{A}(\chi_\eta)\,\overline{B}(\chi_K) \, \Big(  - 1/3\,\chi_K \Big)
\nonumber\\&&
       + \overline{A}(\chi_\eta)\,\overline{B}(\chi_\eta) \, \Big(  - 7/108\,\chi_\pi + 4/27\,\chi_K \Big)
\nonumber\\&&
       + \overline{A}(\chi_\pi) \, \Big(  - 36\,\chi_\pi\,L^r_{4}-
        24\,\chi_\pi\,L^r_{5}+ 72\,\chi_\pi\,L^r_{6}+ 48\,\chi_\pi\,L^r_{8}
        - 24\,\chi_K\,L^r_{4}+ 48\,\chi_K\,L^r_{6}\Big)
\nonumber\\&&
       + \overline{A}(\chi_K) \, \Big(  - 8\,\chi_\pi\,L^r_{4}
        + 16\,\chi_\pi\,L^r_{6}- 48\,\chi_K\,L^r_{4}- 16\,\chi_K\,L^r_{5}
       + 96\,\chi_K\,L^r_{6}+ 32\,\chi_K\,L^r_{8}\Big)
\nonumber\\&&
       + \overline{A}(\chi_\eta) \, \Big( 4/3\,\chi_\pi\,L^r_{4}+ 8/9\,\chi_\pi\,L^r_{5}- 8/3\,\chi_\pi\,L^r_{6}+ 64/3\,\chi_\pi\,L^r_{7}+
         16/3\,\chi_\pi\,L^r_{8}
\nonumber\\&&
 \qquad   - 40/3\,\chi_K\,L^r_{4}
  - 32/9\,\chi_K\,L^r_{5}+ 80/3\,\chi_K\,L^r_{6}- 64/3\,\chi_K\,
         L^r_{7}\Big)
\nonumber\\&&
       + \overline{B}(\chi_\pi) \, \Big(  - 24\,\chi_\pi\,\chi_K\,L^r_{4}+ 48\,\chi_\pi\,\chi_K\,L^r_{6}- 12\,\chi_\pi^2\,L^r_{4}
          - 12\,\chi_\pi^2\,L^r_{5}+ 24\,\chi_\pi^2\,L^r_{6}+ 24\,\chi_\pi^2\,L^r_{8}\Big)
\nonumber\\&&
       + \overline{B}(\chi_K) \, \Big(  - 8\,\chi_\pi\,\chi_K\,L^r_{4}+ 16\,\chi_\pi\,\chi_K\,L^r_{6}- 16\,\chi_K^2\,L^r_{4}-
         8\,\chi_K^2\,L^r_{5}+ 32\,\chi_K^2\,L^r_{6}+ 16\,\chi_K^2\,L^r_{8}\Big)
\nonumber\\&&
       + \overline{B}(\chi_\eta) \, \Big(  - 8/9\,\chi_\pi\,\chi_K\,L^r_{4}+ 32/27\,\chi_\pi\,\chi_K\,L^r_{5}+ 16/9\,\chi_\pi\,
         \chi_K\,L^r_{6}- 128/9\,\chi_\pi\,\chi_K\,L^r_{7}
\nonumber\\&&
 \qquad - 64/9\,\chi_\pi\,\chi_K\,L^r_{8}+ 4/9\,\chi_\pi^2\,L^r_{4}- 4/27
         \,\chi_\pi^2\,L^r_{5}- 8/9\,\chi_\pi^2\,L^r_{6}+ 64/9\,\chi_\pi^2\,L^r_{7}+ 8/3\,\chi_\pi^2\,L^r_{8}
\nonumber\\&&
 \qquad - 32/9\,
         \chi_K^2\,L^r_{4}- 64/27\,\chi_K^2\,L^r_{5}+ 64/9\,\chi_K^2\,L^r_{6}+ 64/9\,\chi_K^2\,L^r_{7}+ 64/9\,
         \chi_K^2\,L^r_{8}\Big)
\nonumber\\&&
 \qquad      + 192\,\chi_\pi\,\chi_K\,C^r_{21} + 8\,\chi_\pi\,\chi_K\,C^r_{94} + 48\,\chi_\pi^2\,C^r_{19} + 80\,\chi_\pi^2\,C^r_{20} +
         48\,\chi_\pi^2\,C^r_{21}
\nonumber\\&&
 - 4\,\chi_\pi^2\,C^r_{94} + 64\,\chi_K^2\,C^r_{20}
  + 192\,\chi_K^2\,C^r_{21}\,,
\nonumber\\
v^s_6 &=&
       + \overline{A}(\chi_\pi)\,\overline{B}(\chi_\eta) \, \Big( 1/3\,\chi_\pi \Big)
       + \overline{A}(\chi_K)\,\overline{A}(\chi_\eta) \, \Big(  - 2/3 \Big)
       + \overline{A}(\chi_K)\,\overline{B}(\chi_\eta) \, \Big(  - 8/9\,\chi_K \Big)
\nonumber\\&&
       + \overline{A}(\chi_\eta)^2 \, \Big( 2/9 \Big)
       + \overline{A}(\chi_\eta)\,\overline{B}(\chi_K) \, \Big(  - 2/3\,\chi_K \Big)
       + \overline{A}(\chi_\eta)\,\overline{B}(\chi_\eta) \, \Big(  - 7/27\,\chi_\pi + 16/27\,\chi_K \Big)
\nonumber\\&&
       + \overline{A}(\chi_\pi) \, \Big(  - 24\,\chi_\pi\,L^r_{4}+ 48\,\chi_\pi\,L^r_{6}\Big)
\nonumber\\&&
       + \overline{A}(\chi_K) \, \Big(  - 16\,\chi_\pi\,L^r_{4}+ 32\,\chi_\pi\,L^r_{6}- 64\,\chi_K\,L^r_{4}- 32\,\chi_K\,L^r_{5}+
         128\,\chi_K\,L^r_{6}+ 64\,\chi_K\,L^r_{8}\Big)
\nonumber\\&&
       + \overline{A}(\chi_\eta) \, \Big(  - 8/3\,\chi_\pi\,L^r_{4}+ 32/9\,\chi_\pi\,L^r_{5}+ 16/3\,\chi_\pi\,L^r_{6}- 128/3\,\chi_\pi\,
         L^r_{7}- 64/3\,\chi_\pi\,L^r_{8}
\nonumber\\&&
  - 64/3\,\chi_K\,L^r_{4}
- 128/9\,\chi_K\,L^r_{5}+ 128/3\,\chi_K\,L^r_{6}+
         128/3\,\chi_K\,L^r_{7}+ 128/3\,\chi_K\,L^r_{8}\Big)
\nonumber\\&&
       + \overline{B}(\chi_K) \, \Big(  - 16\,\chi_\pi\,\chi_K\,L^r_{4}+ 32\,\chi_\pi\,\chi_K\,L^r_{6}- 32\,\chi_K^2\,L^r_{4}
          - 16\,\chi_K^2\,L^r_{5}+ 64\,\chi_K^2\,L^r_{6}
\nonumber\\&&
 \qquad + 32\,\chi_K^2\,L^r_{8}\Big)
\nonumber\\&&
       + \overline{B}(\chi_\eta) \, \Big(  - 32/9\,\chi_\pi\,\chi_K\,L^r_{4}+ 128/27\,\chi_\pi\,\chi_K\,L^r_{5}+ 64/9\,\chi_\pi
         \,\chi_K\,L^r_{6}- 512/9\,\chi_\pi\,\chi_K\,L^r_{7}
\nonumber\\&&
\qquad - 256/9\,\chi_\pi\,\chi_K\,L^r_{8}+ 16/9\,\chi_\pi^2\,L^r_{4}-
         16/27\,\chi_\pi^2\,L^r_{5}- 32/9\,\chi_\pi^2\,L^r_{6}+ 256/9\,\chi_\pi^2\,L^r_{7}
\nonumber\\&&
\qquad + 32/3\,\chi_\pi^2\,L^r_{8}
          - 128/9\,\chi_K^2\,L^r_{4}- 256/27\,\chi_K^2\,L^r_{5}+ 256/9\,\chi_K^2\,L^r_{6}+ 256/9\,\chi_K^2
         \,L^r_{7}
\nonumber\\&&
\qquad + 256/9\,\chi_K^2\,L^r_{8}\Big)
\nonumber\\&&
       - 192\,\chi_\pi\,\chi_K\,C^r_{19} - 64\,\chi_\pi\,\chi_K\,C^r_{20} + 192\,\chi_\pi\,\chi_K\,C^r_{21} + 48\,\chi_\pi^2\,C^r_{19}
          + 16\,\chi_\pi^2\,C^r_{20}
\nonumber\\&&
 + 48\,\chi_\pi^2\,C^r_{21}
 + 4\,\chi_\pi^2\,C^r_{94} + 192\,\chi_K^2\,C^r_{19} +
         192\,\chi_K^2\,C^r_{20} + 192\,\chi_K^2\,C^r_{21}\,.
\ea
These results agree analytically with those of Ref.~\cite{ABT1}.
In Ref.~\cite{ABT1} numerical results were presented. Using the formulas
above one obtains much smaller numerical corrections at NNLO then were
obtained there. This effect is mainly due to the rewriting of the $1/F_0^2$
into $1/F_\pi^2$ and to a lesser extent of rewriting the masses in terms
of the physical masses. 

\begin{figure}
\begin{minipage}{0.48\textwidth}
\includegraphics[width=\textwidth]{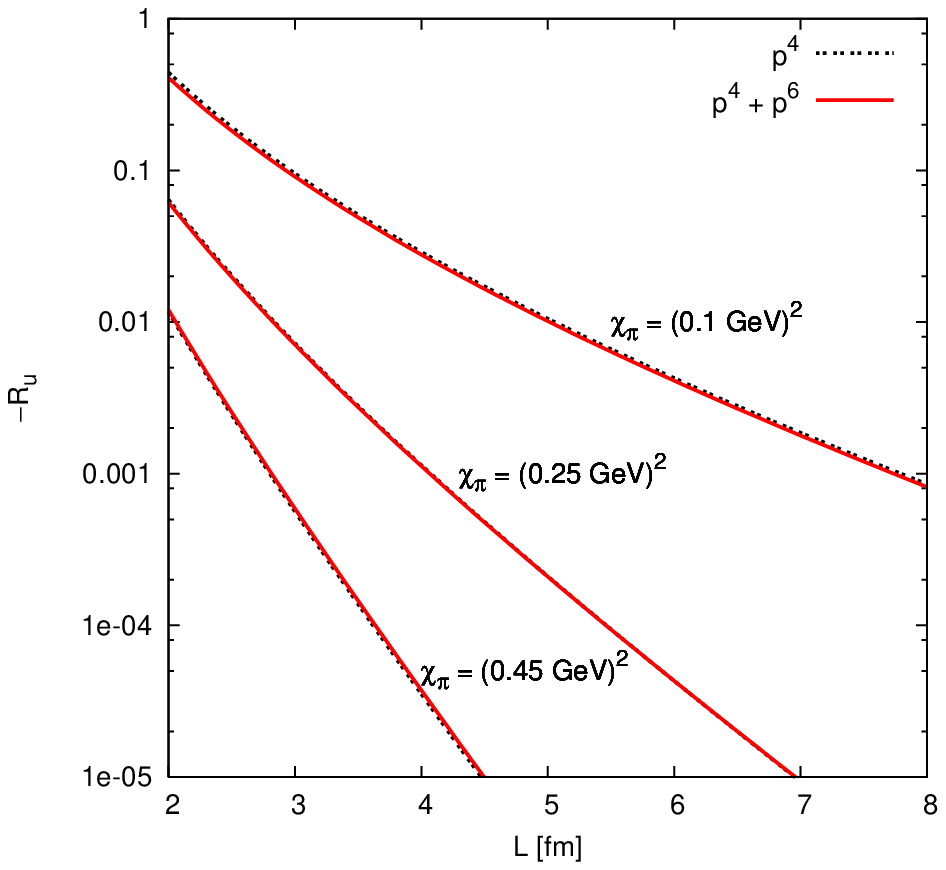}
\centerline{(a)}
\end{minipage}
\begin{minipage}{0.48\textwidth}
\includegraphics[width=\textwidth]{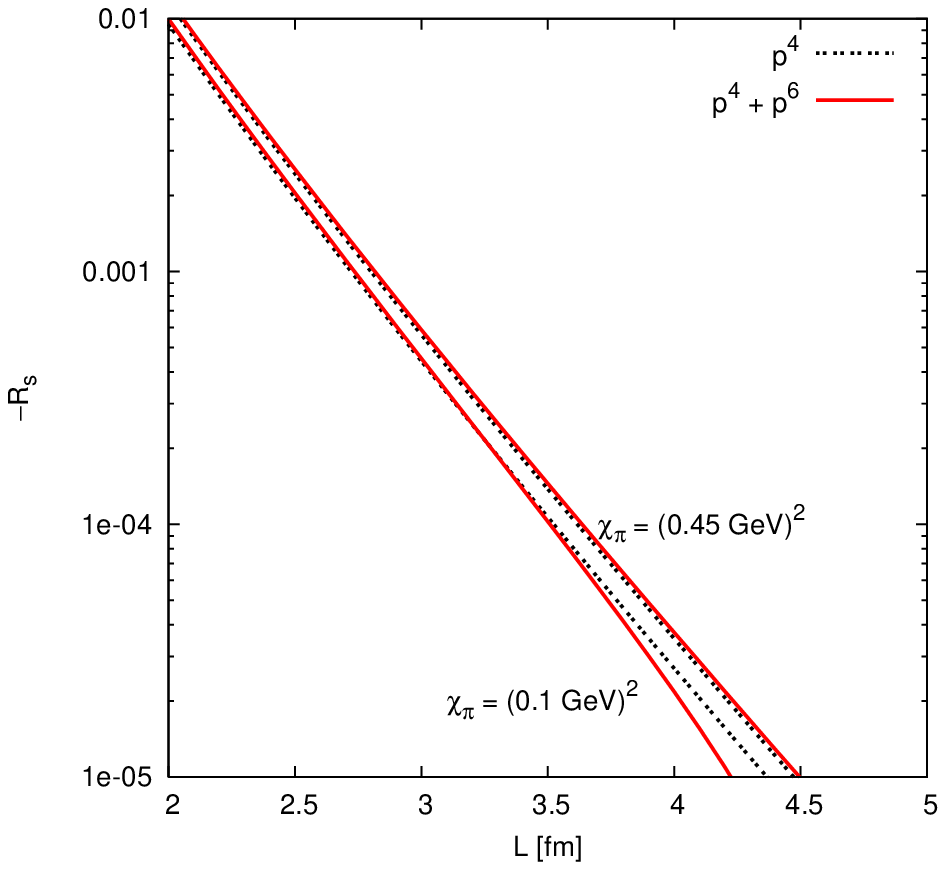}
\centerline{(b)}
\end{minipage}
\caption{\label{figRi}
(a) The ratio $R_u$ for three values of the input lowest order pion mass
and $\chi_K=(450~$MeV$)^2$
(b) The same for $R_s$, with $\chi_K=(450~$MeV$)^2$
and  $\chi_\pi=(100~$MeV$)^2$, 
and $(450~$MeV$)^2$.}
\end{figure}

Numerical results are presented in terms of the ratio
\be
R_q = 
\frac{\dsp\langle\overline qq\rangle_V-\langle\overline qq\rangle_\infty}
{\langle\overline qq\rangle_\infty}
\ee
where we calculate both numerator and denominator to NLO or NNLO in ChPT.
As input parameters for $F_0$ and the $L_i^r$ we use the values
obtained in fit 10 of Ref.~\cite{ABT3}. In addition, we have set $H_2^r=0$.
The results for both $R_u$ and $R_s$
are shown in Fig.~\ref{figRi} for $\chi_K=(450~$MeV$)^2$ and three
values of the lowest order pion mass $\chi_\pi=(100~$MeV$)^2$, $(250~$MeV$)^2$
and $(450~$MeV$)^2$. The finite volume corrections to the strange quark 
vacuum expectation value are always small. The light quark
vacuum expectation value can have sizable finite volume corrections
for the smaller pion mass.

\section{Comparison and conclusions}

The sigma terms in infinite volume are known to two-loop order in ChPT. 
Either directly \cite{BD} or via the derivative of the meson mass to
the quark mass \cite{ABT1,ABT2}. The sigma terms
are also known for two-flavour case \cite{Buergi,pipi,BCT}. 

A major motivation for this work was to test the accuracy of the L\"uscher
type of finite volume formulas. So, how well do the two approaches
compare.

At one-loop the comparison is rather trivial since at most one-propagator
can show up in the relevant one-loop diagrams. We can also apply the
formula (\ref{resultluscher}) for the different species of meson
separately and thus construct the exact one loop ChPT formula. At one loop
the only test possible is thus how fast the sum over $k$ converges
and how quickly it converges to an exponential.
This was already studied in \cite{CD} and the convergence to the leading
exponential is rather slow while the sum over $k$ converges
faster. We find the same conclusions. The extended L\"uscher
formula agrees analytically with the full one-loop expressions and is thus
fully accurate.

At two loop level the two formulas have a different behaviour. There
are diagrams now allowing two propagators simultaneously to feel the
effect of the finite volume. A small complication that needs to be taken
into account here is that the L\"uscher formula is
with the infinite volume mass
at one loop. Thus when changing from the lowest order mass to the physical
mass in the one loop formulas this needs to be done with the infinite volume
expressions. But, even after doing this, the corrections are very small.
The two-loop calculation as plotted in Fig.~\ref{figRi} is obviously very
small. In addition, it is dependent on precisely how one defines the
one-loop order. E.g., there are ambiguities in using the
the physical pion decay constant or the lowest order one, and
how much one uses the Gell-Mann-Okubo relation in the one-loop expression.
With these changes the two-loop calculation can be changed significantly
but remains mostly small. The L\"uscher formula at this order has also
small corrections since the sigma terms have small corrections at one-loop.
We have not plotted it since it will essentially be on top of the other curves
in Fig.~\ref{figRi}.

Both the extended L\"uscher formula and the full two-loop calculations
thus indicate small two-loop order corrections. The actual numerical results
of the two-loop expressions depends strongly on the inherent ambiguities
in defining it. So, both approaches give comparable results at this order,
but we cannot draw conclusions on the accuracy of the extended L\"uscher
formula. 

The extended L\"uscher formula allows also to include effects from even higher
orders by using the sigma terms at higher orders. This quantity is
known to NNLO but its numerical value depends strongly on
the input parameters chosen~\cite{BD}. There might thus be sizable effects
at higher orders but there need not be.

In conclusion, we have derived an extended L\"uscher formula for the
finite volume effects on the quark vacuum condensate. We have also calculated
these effects to two-loop order in ChPT. At one-loop order the extended
L\"uscher formula is exactly equal to the full ChPT calculation.
At two-loop order, the latter includes extra effects, but both approaches
indicate very small corrections. The difference is within the inherent
uncertainty of the full ChPT calculation at that order and we thus cannot
conclude if the difference is due to the inherent uncertainty in the
two-loop order result or to the effects not captured by the L\"uscher formula.

\section*{Acknowledgements}
We thank Gilberto Colangelo and Christoph Haefeli for discussions. 
This work is supported by the European Union TMR network,
Contract No. 
HPRN-CT-2002-00311  (EURIDICE) and by 
the European Community-Research Infrastructure
Activity Contract No. RII3-CT-2004-506078 (HadronPhysics).
KG acknowledges a fellowship from the Iranian Ministry of Science.

\end{document}